\documentclass[prd,reprint,amsmath,nofootinbib,superscriptaddress,showpacs]{revtex4-1}

\usepackage[T1]{fontenc}
\usepackage[utf8]{inputenc}
\usepackage[brazil,english]{babel}
\usepackage{tensor}
\usepackage{amssymb}
\usepackage[usenames,dvipsnames]{color}
  \definecolor{darkblue}{RGB}{0,0,150}
\usepackage[unicode]{hyperref}
\usepackage{nicefrac}
\usepackage{graphicx}
\usepackage{mathrsfs}
\usepackage{subcaption}
\hypersetup{
  pdfinfo = {
    Author = {Niayesh Afshordi, Michele Fontanini, Daniel C. Guariento},
    Title = {Horndeski meets McVittie: A scalar field theory for accretion onto cosmological black holes}
  },
  colorlinks = true,
  breaklinks = true,
  linkcolor = darkblue,
  urlcolor = darkblue,
  citecolor = darkblue
}

\newcommand{\ud}{\ensuremath{\mathrm{d}}}
\usepackage{soul}

\begin{document}

\title{Horndeski meets McVittie:\\ A scalar field theory for accretion onto cosmological black holes}

\author{Niayesh Afshordi}
\email{nafshordi@pitp.ca}

\affiliation{Department of Physics \& Astronomy, University of Waterloo, Waterloo, ON, N2L 3G1, Canada}

\affiliation{Perimeter Institute for Theoretical Physics, 31 Caroline St. N., Waterloo, ON, N2L 2Y5, Canada}

\author{Michele Fontanini}
\email{fmichele@fma.if.usp.br}

\affiliation{\foreignlanguage{brazil}{Instituto de Física, Universidade de São Paulo, Caixa Postal 66.318, 05315-970, São Paulo, SP}, Brazil}

\author{Daniel C. Guariento}
\email{carrasco@fma.if.usp.br}
\email{dguariento@pitp.ca}

\affiliation{\foreignlanguage{brazil}{Instituto de Física, Universidade de São Paulo, Caixa Postal 66.318, 05315-970, São Paulo, SP}, Brazil}

\affiliation{Perimeter Institute for Theoretical Physics, 31 Caroline St. N., Waterloo, ON, N2L 2Y5, Canada}

\begin{abstract}

We show that the generalized McVittie spacetime, which represents a black hole with time-dependent mass in an expanding universe, is an exact solution of a subclass of the Horndeski family of actions. The heat-flow term responsible for the energy transfer between the black hole and the cosmological background is generated by the higher-order kinetic gravity braiding term, which generalizes the cuscuton action that yields McVittie with constant mass as a solution. Finally, we show that this generalization can be understood in terms of a duality realized by a disformal transformation, connecting the cuscuton field theory to an extension of the Horndeski action which does not propagate any scalar degrees of freedom. Our finding opens a novel window into studies of non-trivial interactions between dark energy/modified gravity theories and astrophysical black holes.

\end{abstract}


\maketitle

\section{Introduction}

Until recently, the question of whether cosmic evolution affects local gravitationally bound systems was seldom addressed from the point of view of exact solutions of general relativity. However, a renewed interest in such non-vacuum (and in particular black-hole) solutions has recently been raised, fueled by numerous exciting developments in cosmology and astrophysics, such as non-trivial interactions of scalar fields  with astrophysical black holes \cite{Davis:2014tea,Cardoso:2013fwa}, effects of modified gravity on stellar evolution \cite{Sakstein:2013pda}, constraints from Planck satellite observations on different scalar-field inflationary models for structure formation and subsequent cosmological evolution \cite{Martin:2013nzq} and, from a formal point of view, loopholes in the no-hair theorem \cite{Sotiriou:2013qea,*Sotiriou:2014pfa} which can be exploited by these theories. Recently, subsets of the Horndeski action \cite{Horndeski:1974wa} have been considered as candidates for black hole hair, and exact stationary solutions were found for some special cases \cite{Babichev:2013cya,*Anabalon:2013oea,*Kobayashi:2014eva,*Charmousis:2014zaa}, as well as non-stationary perturbed solutions \cite{Herdeiro:2014goa,*Herdeiro:2014ima}. Most interestingly, it has been proposed that dark energy might be the quantum hair of astrophysical black holes, which could potentially solve the cosmic coincidence (i.e. ``why now?'') problem \cite{PrescodWeinstein:2009mp,Afshordi:2010eq}. Nevertheless, the details of the influence that cosmic expansion has on the local causal structure of black holes have remained an elusive subject, but for few notable exceptions (e.g. \cite{Firouzjaee:2014zfa}).

Among solutions of general relativity that describe black holes evolving in time, the McVittie solution \cite{McVittie:1933zz,*mcvittie-1932} is certainly one of the most ubiquitous, appearing on a wide range of problems stemming from perfect-fluid cosmology to scalar-field actions and modified theories of gravity \cite{Faraoni:2013aba}. It is the unique perfect-fluid solution of Einstein's equations to display a central singularity in a spherically symmetric shear-free metric that is asymptotically FLRW for large distances \cite{raychaudhuri}. Moreover, it has been shown that, when treated as a perturbation of an asymptotically Schwarzschild-de~Sitter metric, strong evidence exists pointing towards its stability \cite{Chadburn:2013mta}. The McVittie metric is also the oldest member of the physically interesting Kustaanheimo-Qvist class of solutions \cite{stephani-exact,*kustaanheimo-1947}, and in the decades following its original derivation, its apparent simplicity and non-trivial causal structure have spawned much discussion in the literature \cite{Nolan:1998xs,*Nolan:1999kk,*Nolan:1999wf,Faraoni:2012gz}. In fact, an almost 80-year-long debate has continued until very recently when the conditions under which this spacetime actually describes a black hole were finally settled \cite{Kaloper:2010ec,Lake:2011ni}, as well as how particle motion and the structure of the event horizon ultimately depend on the cosmological history \cite{daSilva:2012nh,Nolan:2014maa}. Moreover, going past the classical analysis of GR solutions, and entering the realm of field theory, it was recently found \cite{Abdalla:2013ara} that the McVittie metric is an exact solution for an extreme limit of $k$-essence, known as the \emph{cuscuton} theory \cite{Afshordi:2006ad,*Afshordi:2007yx}, which in turn is the unique $k$-essence able to support the McVittie solution. In light of this result and the analogue descriptions that the cuscuton field admits \cite{Afshordi:2009tt}, a natural consequence is that the presence of constant-mean-curvature (CMC) surfaces on the constant-time foliation also makes McVittie a solution of Ho\v{r}ava-Lifshitz gravity with anisotropic Weyl symmetry, and it can also be considered as a non-trivial example of an exact solution in Shape Dynamics \cite{Gomes:2010fh}.

In this work we aim at going further in the description of these dynamic black holes by introducing time evolution of the central mass as a result of interaction with the scalar field source. A seminal step in this direction was taken when a generalized version of the McVittie metric was introduced with an arbitrary time-dependent mass \cite{Faraoni:2007es}, which generalized the Sultana-Dyer black hole \cite{Sultana:2005tp} (in its turn obtained via a conformal transformation of the Schwarzschild metric), and was later applied to the problem of dark-energy accretion \cite{Gao:2008jv,*Faraoni:2008tx}. The specific form of the mass function could then in principle be obtained from the field equations, or from some hydrodynamic model for the source, which no longer could be taken to be a single perfect fluid \cite{Carrera:2009ve} or a superposition of multiple perfect fluids \cite{guariento-tese-2010}, so a more general approach was needed. The needed step forward arrived when it was found that in the fluid interpretation the accretion (or evaporation) rate could be explained by the introduction of a heat-flow term associated to the presence of a temperature gradient via a Landau-Eckart model \cite{Guariento:2012ri}, as other viscous terms vanish due to the high degree of symmetry of the metric.

In light of these results, the natural question which arises is whether the generalized McVittie metric \cite{Guariento:2012ri} can be derived from an action principle. An affirmative answer means that it is more than simply a naive generalization of the original McVittie solution, rather that there is a way to describe the interaction between a black hole and a cosmological background in more fundamental terms, and, more importantly, that by studying such a simple analytic solution we can gain insight into basic aspects of black holes with scalar hair, which are quickly becoming common-place results of many modern theories.

The answer to this question is indeed affirmative. By considering the Horndeski action, the most general self-gravitating scalar field action that gives second-order equations of motion \cite{Horndeski:1974wa}, we demonstrate that the generalized McVittie metric can be obtained as an exact solution. We use the symmetries of the metric to constrain the coefficients of the full Horndeski action, and cast the equations of motion in terms of these general functions. Interestingly, the relation between the free functions that define a particular generalized McVittie solution, namely the mass and the cosmological scale factor, and the field contains a certain amount of indeterminacy. This is not surprising, since the same behavior was observed when considering solutions with a cuscuton source, suggesting that this particular form of the Horndeski action is a generalization of the non-dynamic ``parasitic'' nature \cite{Afshordi:2006ad} of its $k$-essence counterpart.

The paper is organized as follows: in Sec.\ \ref{sec:horndeski} we review the basic aspects of the Horndeski scalar field and its analogous fluid description; in Sec.\ \ref{sec:mcvittie-gen} we show that the generalized McVittie metric can be obtained as an exact solution to a particular case of the theory (namely, the kinetic gravity braiding sector), and we present a particular complete solution to the system of equations; in Sec.\ \ref{sec:cuscuton} we discuss the dynamical properties of the field in parallel with the cuscuton theory. We present our conclusions in Sec.\ \ref{sec:conclusions}. Throughout the paper, Greek indices run from 0 to 3, Latin indices run from 1 to 3 and we use the $(-,+,+,+)$ signature. Time derivatives are denoted with an overhead dot.

\section{Higher-order actions}\label{sec:horndeski}

The Horndeski scalar action \cite{Horndeski:1974wa,Gao:2011qe,*Kobayashi:2011nu} can be cast in the following form:

\begin{equation}\label{Shorn-full}
  S = \int \ud^4 x \sqrt{-g} \left( \frac{1}{2} R + \sum_{n = 2}^{5} \mathcal{L}^{(n)} \right) \,,  
\end{equation}
where
\begin{align}
  \mathcal{L}^{(2)} =\,& G^{(2)} (X, \varphi) \,,\label{kessence-part}\\
  \mathcal{L}^{(3)} =\,& G^{(3)} (X, \varphi)\, \square \varphi \,,\label{kgb-part}\\
\begin{split}
  \mathcal{L}^{(4)} =\,& G^{(4)}_{,X}(X, \varphi) \left[ \left( \square \varphi \right)^2 - \nabla_{\alpha} \nabla_{\beta} \varphi \, \nabla^{\alpha} \nabla^{\beta} \varphi \right] \\
  & + R \, G^{(4)}(X, \varphi) \,,
\end{split}\\
\begin{split}
\mathcal{L}^{(5)} =\,& G^{(5)}_{,X}(X, \varphi) \left[ \left( \square \varphi \right)^3 - 3 \square \varphi \, \nabla_{\alpha} \nabla_{\beta} \varphi \, \nabla^{\alpha} \nabla^{\beta} \varphi \right. \\
  & \left.\vphantom{\left( \square \varphi \right)^3} + 2 \nabla_{\alpha} \nabla_{\beta} \varphi \, \nabla^{\alpha} \nabla^{\rho} \varphi \, \nabla_{\rho} \nabla^{\beta} \varphi \right] \\
  & - 6 G_{\mu\nu} \nabla^{\mu} \nabla^{\nu} \varphi \, G^{(5)}(X, \varphi) \,,
\end{split}
\end{align}
and where
\begin{equation} \label{Xdef}
  X \equiv -\frac{1}{2} g^{\alpha\beta} \nabla_{\alpha} \varphi \nabla_{\beta} \varphi \,,
\end{equation}
is the canonical kinetic term, $R$ is the Ricci scalar, $G_{\mu\nu}$ is the Einstein tensor and $\square \varphi = g^{\alpha \beta} \nabla_\alpha \nabla_\beta \varphi$. The $G^{(n)}$ are arbitrary functions of the field and kinetic term. We denote by $\nabla_{\alpha}$ the covariant derivative compatible with the metric $g_{\mu\nu}$, and we also use the notation $G^{(n)}_{,X} = \partial_X G^{(n)}$ and $G^{(n)}_{,\varphi} = \partial_{\varphi} G^{(n)}$. It is the most general single scalar field action one can write containing higher derivatives of the field and still yielding second-order equations of motion\footnote{This class of theories can be expanded by allowing for higher-order equations of motion which, through gauge transformations, lead to propagating degrees of freedom that follow second-order dynamics \cite{Gleyzes:2014dya}.}. Some interesting discussions have appeared recently in the literature regarding a possible loophole in this claim, in which more general terms could in principle be considered \cite{Zumalacarregui:2013pma,*Zumalacarregui:2012us}, and also the subcases in which the action \eqref{Shorn-full}, taken as a generalization of scalar-tensor theories, can be cast in the Einstein frame \cite{Bettoni:2013diz}.

If we vary the action \eqref{Shorn-full} with respect to the metric, the corresponding energy-momentum tensor can be interpreted in terms of projections onto the comoving flow. In that interpretation, the contribution from the term $\mathcal{L}^{(2)}$ always takes the form of a perfect fluid, the term $\mathcal{L}^{(3)}$ modifies the perfect-fluid components from the previous term and also gives rise to a component associated with heat flow, and the terms $\mathcal{L}^{(4)}$ and $\mathcal{L}^{(5)}$ will also give rise to more complicated dissipative terms.

Since we are interested in shear-free solutions, we consider for simplicity a subcase of action \eqref{Shorn-full}, in which we only include the first additional term $\mathcal{L}^{(3)}$ to the $k$-essence action $\mathcal{L}^{(2)}$, producing an action known in the literature as \emph{kinetic gravity braiding} \cite{Deffayet:2010qz,Pujolas:2011he,babichev-2012}. It can be shown that this is the most general term whose resulting energy-momentum tensor has vanishing anisotropic stress in the comoving frame for generic forms of the function coefficients. We therefore reduce the purely scalar sector of the general action \eqref{Shorn-full} to
\begin{equation} \label{Shorn}
  S_{\varphi} = \int \ud^4 x \sqrt{-g} \left[ \mathcal{L}^{(2)} + \mathcal{L}^{(3)} \right] .
\end{equation}
When the field is minimally coupled to gravity the full action we consider is
\begin{equation}\label{SKGB}
  S = \int \ud^4 x \sqrt{-g} \left[ \frac{1}{2} R + G^{(2)} (X, \varphi) + G^{(3)} (X, \varphi) \square \varphi \right] .
\end{equation}

The energy-momentum tensor which results from varying the scalar sector of the action \eqref{SKGB} with respect to the metric is
\begin{equation}
\begin{split}
  T_{\mu\nu} =&\, - \frac{2}{\sqrt{-g}} \frac{\delta S_{\varphi}}{\delta g^{\mu\nu}} \\
  =&\, \left( G^{(2)} - \nabla_{\alpha} G^{(3)} \, \nabla^{\alpha} \varphi \right) g_{\mu\nu} + 2 \nabla_{\left( \mu \right.} G^{(3)} \, \nabla_{\left. \nu \right)} \varphi \\
  & + \left( G^{(2)}_{,X} + \square \varphi G^{(3)}_{,X} \right) \nabla_{\mu} \varphi \nabla_{\nu} \varphi \,. \label{tmn-horn}
\end{split}
\end{equation}
The fluid analogy is then carried out by defining the equivalent fluid flow $u^{\mu}_{(\varphi)}$ as
\begin{equation} \label{4vel}
  u^{\mu}_{(\varphi)} \equiv \frac{\nabla^{\mu} \varphi}{\sqrt{2 X}} \,,
\end{equation}
where the denominator is chosen to normalize the four-velocity, so $u_{\mu} u^{\mu} = -1$. We also define the orthogonal projector with respect to the flow as
\begin{equation}\label{3metric}
\begin{split}
  \gamma_{\mu\nu} &\equiv g_{\mu\nu} + u_{\mu} u_{\nu} \\
  &= g_{\mu\nu} + \frac{\nabla_{\mu} \varphi \, \nabla_{\nu}\varphi}{2 X} \,.
\end{split}
\end{equation}
The decomposition of the energy-momentum tensor \eqref{tmn-horn} in quantities relative to the flow defined by \eqref{4vel}, may be interpreted as the functional definition of the equivalent fluid quantities \cite{Pujolas:2011he}
\begin{align}
  \begin{split}
    \rho \equiv&\, - \left( G^{(2)} - \nabla_{\alpha} G^{(3)} \, \nabla^{\alpha} \varphi \right) \\
    &\, + 2 X \left( G^{(2)}_{,X} + \square \varphi \, G^{(3)}_{,X} \right) - 2 \nabla_{\alpha} G^{(3)} \, \nabla^{\alpha} \varphi \,,
  \end{split}\\
  p \equiv&\, G^{(2)} - \nabla_{\alpha} G^{(3)} \, \nabla^{\alpha} \varphi \,,\\
  \begin{split}  
    q^{\mu} \equiv&\, \sqrt{2 X} \left( \nabla^{\mu} G^{(3)} + \frac{1}{2X} \nabla_{\alpha} G^{(3)} \, \nabla^{\alpha} \varphi \nabla^{\mu} \varphi \right)\\
    =&\, \sqrt{2 X} \tensor{\gamma}{^\mu _\alpha} \nabla^{\alpha} G^{(3)} \,,
  \end{split}
\end{align}
where $\rho$ can be seen as the energy density, $p$ as the pressure and $q^{\mu}$ as the heat flow vector.

\section{McVittie metric with time-dependent mass}\label{sec:mcvittie-gen}

The McVittie metric \cite{McVittie:1933zz,*mcvittie-1932}
\begin{equation} \label{mcvittie}
\begin{split}
  \ud s^2 =&\, -\left( \frac{1 - \frac{m}{2 a r}}{1 + \frac{m}{2 a r}} \right)^2 \ud t^2 \\
  & + a^2 \left( 1 + \frac{m}{2 a r} \right)^4 \left( \ud r^2 + r^2 \ud \Omega^2_2 \right) \,,
\end{split}
\end{equation}
where $m$ is a constant and $a = a(t)$ is a classical perfect-fluid solution to Einstein's equations and has been recently shown to be also an exact solution of a system with a self-gravitating scalar field with a modified kinetic term (quadratic cuscuton) \cite{Abdalla:2013ara}.

\begin{figure}[!htp]
  \centering
  \begin{subfigure}{.5\textwidth}
    \includegraphics[width=\textwidth]{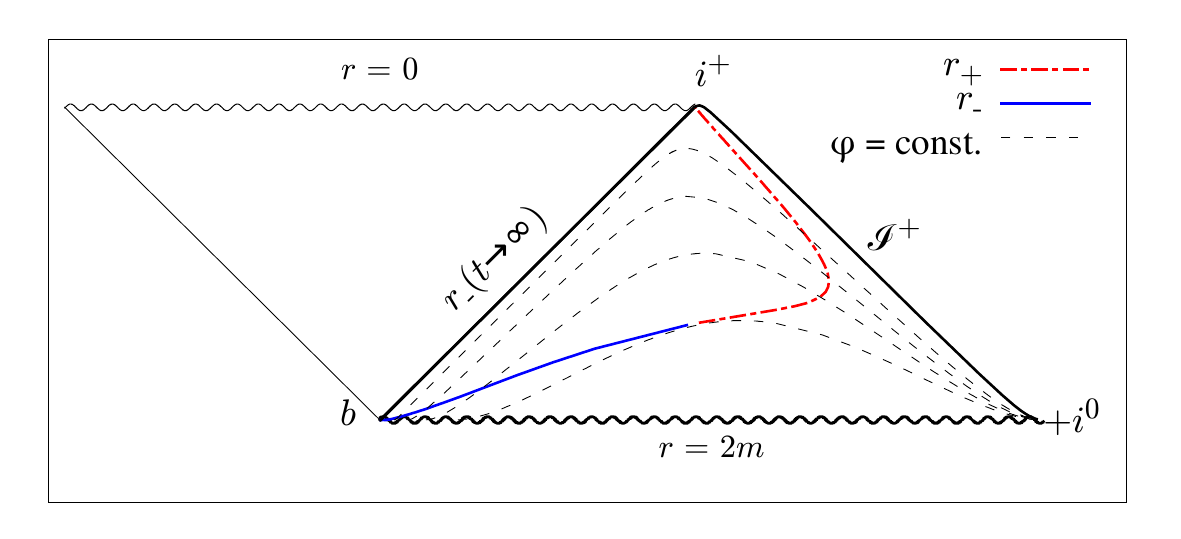}
    \caption{McVittie metric \eqref{mcvittie} with constant mass}
  \end{subfigure}
  \begin{subfigure}{.5\textwidth}
    \includegraphics[width=\textwidth]{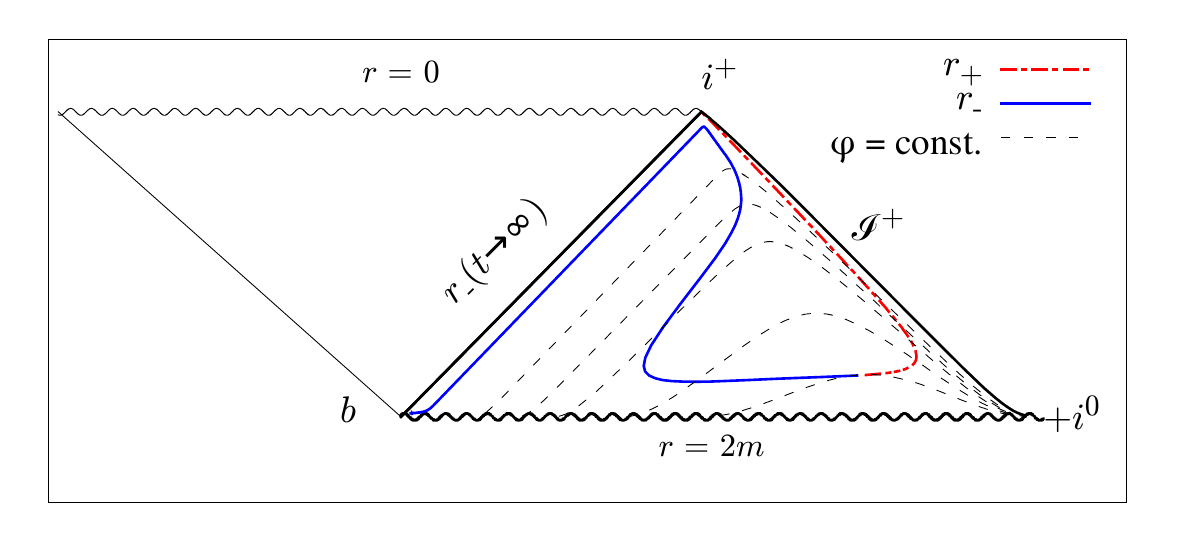}
    \caption{Generalized McVittie with increasing mass. The mass function is of the form $m (t) = m_0 \left[ \frac{1}{2} + \tanh \left( t - t_0 \right) \right]$.}
  \end{subfigure}
  \caption{Examples of the causal structure of the McVittie and generalized McVittie metrics, depicting the apparent horizons and surfaces of constant time. A Schwarzschild-de~Sitter extension is patched at the left of each diagram. In both examples, the expansion is given by Eq.\ \eqref{Hlcdm}. Note that, in this class of metrics, the null infinity $\mathscr{I}^+$ is space-like. }
  \label{fig:mcv}
\end{figure}

The generalized McVittie metric \cite{Faraoni:2007es} with $m = m(t)$ is known to be an exact solution for a comoving shear-free imperfect fluid with heat flow \cite{Guariento:2012ri}. In generalized McVittie, the kinetic term, the velocity and the expansion read
\begin{subequations}
\begin{align}
  X = & - \frac{1}{2} \left( \frac{2 a(t) r + m(t)}{2 a(t) r -m(t)} \right)^2 \dot{\varphi}^2 \label{expansion-def}\\
  u^\mu = &  \frac{2 a r + m}{2 a r - m}  \delta^\mu_0 \label{Xdef-mcv}\\
  \begin{split}
    \Theta \equiv & \nabla_{\alpha} u^{\alpha} \\
    =& 3 \left( \frac{\dot{a}}{a} + \frac{2 \dot{m}}{2 a r - m} \right) \,, 
  \end{split}
\end{align}
\end{subequations}
corresponding to a future-oriented flow, that in turn using \eqref{4vel} requires the choice $\dot{\varphi}(t)<0$ in the case an homogeneous field is considered, as we will see in the following sections. The Penrose diagrams for particular cases of the McVittie metric \eqref{mcvittie} and its generalization can be seen in Fig.\ \ref{fig:mcv}.

\subsection{Einstein equations}

We now consider the source of the gravitational field to be the scalar described by the action \eqref{Shorn}. The property of spatial Ricci-isotropy of the generalized McVittie solution \cite{Carrera:2009ve} means that the pressure of the source is isotropic, that is, $\tensor{T}{^r_r} = \tensor{T}{^\theta_\theta}$. By examining equation \eqref{tmn-horn}, it can be seen that for non-trivial functions $G^{(2)}$ and $G^{(3)}$ the isotropy condition is only satisfied if the field is homogeneous. We focus therefore on a homogeneous field configuration $\varphi (t)$, which by Eq.\ \eqref{4vel} corresponds to a comoving equivalent fluid. The $(t,r), (t,t)$, and $(r,r)$ Einstein equations then respectively read
\begin{subequations}
\begin{align}
  -\frac{\dot{m}}{m \dot{\varphi}} =&\, X G^{(3)}_{,X} \,,\label{EE01} \displaybreak[0]\\
  -\frac{1}{3} \Theta^2 =&\, G^{(2)} - 2 X \left[ G^{(3)}_{,\varphi} + G^{(2)}_{,X} + \sqrt{2 X} G^{(3)}_{,X} \Theta \right] \,, \label{EE00} \displaybreak[0]\\
\begin{split}
  -\frac{1}{3} \left( \Theta^2 + \vphantom{\frac{2 a r + m}{2 a r - m}}\right.&\left. 2 \dot{\Theta} \frac{2 a r + m}{2 a r - m} \right) \\
  =&\, G^{(2)} + 2 X \left\{ G^{(3)}_{,\varphi} + G^{(3)}_{,X} \left[ \ddot{\varphi} \left( \frac{2 a r + m}{2 a r - m} \right)^2 \right. \right.\\
  & \left.\left.+ 4 r \sqrt{2 X} \frac{m \dot{a} - \dot{m} a}{\left( 2 a r - m \right)^2} \right] \right\} \, , \label{EE11}
\end{split}
\end{align}
\end{subequations}
where $\Theta$ and $X$ are given by Eqs.\ \eqref{expansion-def} and \eqref{Xdef-mcv} respectively.

The equations above select one or more sets of functions $\{a(t),m(t)\}$ as solutions to the system described by the action \eqref{SKGB}. It has to be noted, though, that by not assigning a fixed functional form for the functions $G^{(2)}$ and $G^{(3)}$ we are actually looking at a family of possible actions, with a plethora of possible solutions. First, given that the only $r$-dependence in Eq.\ \eqref{EE01} is in $X$, we can ``integrate'' Eq.\ \eqref{EE01} along constant time [and thus constant $\varphi(t)$ and $\dot{\varphi}(t)$] hypersurfaces to get\footnote{Eq.\ \eqref{solg} also corresponds to a constant heat conductivity in the fluid interpretation of the KGB action \cite{Pujolas:2011he}.}
\begin{equation}
  G^{(3)} (X,\varphi) = g_0 (\varphi) \ln X + g_1 (\varphi) \,. \label{solg}
\end{equation}
By doing this, we are effectively trading the $(t,r)$ Einstein equation \eqref{EE01} for a first order differential equation for $\varphi$, namely,
\begin{equation}
  g_0 (\varphi) = - \frac{M}{\dot{\varphi}} \,, \label{solg0}
\end{equation}
[where we have defined the function $M (t) \equiv \nicefrac{\dot{m}}{m}$] reducing the freedom in the choice of $G^{(3)}$ to the freedom in the choice of $g_0(\varphi)$ and $g_1(\varphi)$.

Inserting this ansatz in the $(t,t)$ equation \eqref{EE00} we find
\begin{equation}
\begin{split}
  - \frac{1}{3}\Theta^2 =&\, G^{(2)} - 2 X \left( G^{(2)}_{,X} + g_1' \right) + 2 \sqrt{2 X} \frac{M}{\dot{\varphi}} \Theta\\
  &+ 2 X \ln X \left( \frac{\dot{M}}{\dot{\varphi}^2} - \frac{M \ddot{\varphi}}{\dot{\varphi}^3} \right)  \,,
\end{split}
\end{equation}
where the prime denotes a derivative with respect to the field $\varphi$. In a similar way, ``integrating'' along constant time hypersurfaces equation \eqref{EE00} after using Eq. \eqref{EE01} in it, we obtain a functional form for $G^{(2)}$ that reads
\begin{equation} \label{soll}
\begin{split}
  G^{(2)} =&\, f_1 (\varphi) + f_2 (\varphi) \sqrt{X} \\
  & + 2 X \left[ (2 -\ln X) g_0' - g_1' - 3 g_0^2 \right] \, ,
\end{split}
\end{equation}
where $f_1$ has to satisfy
\begin{equation} \label{solf1}
f_1 (\varphi) = -3 \left( H - M \right)^2 \,, 
\end{equation}
and $f_2$ can be obtained by use of equation \eqref{EE11} when the form for $G^{(2)}$ above is used. This gives
\begin{equation} \label{solf2}
f_2 (\varphi) = \frac{2 \sqrt{2}}{\dot{\varphi}} \left[ \dot{H} - \dot{M} + 3 M \left( H - M \right) \right] \,.
\end{equation}

\subsection{Field equations}

In the previous section we have shown that, via a suitable choice of functions $G^{(3)}$ and $G^{(2)}$ appearing in the Lagrangian, we can reduce the quite complex system of Einstein equations for the generalized McVittie metric to a set of first order differential equations for the sourcing homogeneous field $\varphi(t)$. We are going to see now that the choice we made for the form of the Lagrangian automatically satisfies the field equation for $\varphi(t)$.

The field equation from action \eqref{Shorn} reads
\begin{equation}\label{eqmov-horn}
  \begin{split}
    0 =&\, G^{(2)}_{,\varphi} + G^{(3)}_{,\varphi} \square \varphi \\
    & + \nabla_{\alpha} \left[ \left( G^{(2)}_{,X} + G^{(3)}_{,X} \square \varphi \right) \nabla^{\alpha} \varphi + \nabla^{\alpha} G^{(3)} \right]\,.
  \end{split}
\end{equation}
By inserting the functional forms for $G^{(3)}$ from Eq.\ \eqref{solg} and $G^{(2)}$ from \eqref{soll} with coefficients given by \eqref{solg0}, \eqref{solf1} and \eqref{solf2}, it is easy to show that Eq.\ \eqref{eqmov-horn} is identically satisfied.

We notice at this point that the function $g_1(\varphi)$ plays no role in any of the equations we used, it is completely arbitrary and can be dropped from our analysis in the following. We can therefore claim that any triplet $\left\{ \varphi(t), m(t), a(t) \right\}$ will be a solution describing the system given by an action of the form
\begin{equation}\label{action-solution}
\begin{split}
  S =&\, \int \ud^4 x \sqrt{-g} \left\{  f_1 (\varphi) + f_2 (\varphi) \sqrt{X} \vphantom{\frac{1}{2}}\right.\\
  &+ 2 X \left[ (2 -\ln X) g_0'(\varphi) -  3 g_0(\varphi)^2 \right] \\
  &\left. + \left[ g_0 (\varphi) \ln X  \right] \square \varphi + \frac{1}{2} R \right\} \,,
\end{split}
\end{equation}
when three functions $g_0$, $f_1$, $f_2$ of the field $\varphi$ can be found that satisfy Eqs.\ \eqref{solg0}, \eqref{solf1} and \eqref{solf2}.

\subsection{An example}

We want now to show an explicit example of solutions to the equations we have seen in the previous sections. The aim of course is to find not just any solution; rather, it is to find a choice of $m(t)$ and $a(t)$ for which the generalized McVittie metric describes a black hole \cite{Guariento:2012ri} as it is discussed in Appendix \ref{null-geodesics}. 

Possibly the easiest way to approach the problem is to require the three functions $g_0$, $f_1$ and $f_2$ to be constants. This requires $H$ and $M$ differing by a constant, and $\dot{\varphi}$ being proportional to $M$. Let us then assume
\begin{align}
\dot{\varphi}(t) & = - M(t)  \, , \\
M(t) &= H(t) -H_0 \,,
\end{align}
which immediately implies 
\begin{align}
g_0 &= 1 \, , \\
f_1 &= - 3 H_0^2 \, , \\
f_2 &= 6 \sqrt{2} H_0\, . 
\end{align}
This class of solutions contains for instance a $\Lambda$CDM--like expansion with a decreasing accretion rate given by 
\begin{align}
H(t) &=H_0  \coth(H_0 t) \,, \label{Hlcdm}\\
m(t) &= m_0 \, e^{-H_0 t} \sinh (H_0 t) \, , \\
\varphi(t) &= \varphi_0 + H_0 t - \ln \left[ \sinh (H_0 t) \right] \,.
\end{align}

Abandoning the simplification of constant $g_0$, $f_1$ and $f_2$ is also simple, by still maintaining $M(t) = H(t) -H_0$ and the $\Lambda$CDM--like expansion above, it is in fact easy to build exponential solutions for $\varphi$ and a triplet of functions that satisfy our system.

\section{Field dynamics and Cuscuton duality}\label{sec:cuscuton}

A natural question which arises from the previous analysis is whether this particular case of Horndeski action represents a truly dynamical field, or it can be seen as a higher-order analogue to the cuscuton field, which is known to have the constant-mass McVittie metric as an exact solution \cite{Abdalla:2013ara}.

The first thing that can be noted is that Eqs.\ \eqref{eqmov-horn} for the field $\varphi$ do not lose the second-order time derivatives in the homogeneous limit, which is the interesting limit considered here. In fact, it can be shown \cite{Deffayet:2010qz} that the Lagrangian we are considering here \eqref{Shorn} produces a field equation that can in general be cast in the form
\begin{multline}
  L^{\mu\nu} \nabla_\mu \nabla_\nu \varphi + \left( \nabla_\alpha \nabla_\beta \varphi \right)Q^{\alpha \beta \mu \nu} \left( \nabla_\mu \nabla_\nu \varphi \right) \\
  + \mathcal{E}_\varphi - G^{(3)}_{,X} R^{\mu\nu} \nabla_\mu \varphi  \nabla_\nu \varphi = 0 \,.
\end{multline}
In the homogeneous limit $\varphi \to \varphi(t)$, the object containing quadratic terms in second derivatives, $Q^{\alpha \beta \mu \nu}$, vanishes, leaving only linear terms in $\ddot{\varphi}$ characterized by $L^{\mu \nu}$. In particular, the linear term becomes proportional to
\begin{equation}
G^{(2)}_{,X} - 2 G^{(3)}_{,\varphi} - 2 X G^{(3)}_{,\varphi X} + 2 X G^{(2)}_{,X X} \, .
\end{equation}
It follows then that Lagrangians satisfying
\begin{equation}\label{patho-cond}
 X G^{(3)}_{,\varphi} + f (\varphi) = 2 X G^{(2)}_{,X} - G^{(2)} \,,
\end{equation}
for any choice of the function $f (\varphi)$, lose the second-order time derivatives in the homogeneous field limit, in a much similar way as in the cuscuton system.

We can notice at this point that the Lagrangian we considered cannot take the pathological form above if the generalized McVittie metric is to appear as a solution, as can be easily seen using relations \eqref{solg} and \eqref{soll} in \eqref{patho-cond}. Therefore, the straightforward analogy with the cuscuton field in the McVittie spacetime \cite{Abdalla:2013ara} cannot be carried on in this way.

Before moving to a more detailed analysis of the connection between the cuscuton system and the Horndeski one considered here, note that by manipulating Eq.\ \eqref{solg0}, as well as \eqref{solf1} and \eqref{solf2}, we obtain
\begin{equation}\label{g0mphi}
  g_0\left(\varphi \right) = - \frac{\ud \left( \ln m \right)}{\ud \varphi} \, ,
\end{equation}
which, considering that we imposed a nonzero time derivative for the field $\dot{\varphi} \ne 0$,  allows us to cast $m$ implicitly in terms of $\varphi$:
\begin{equation} \label{mdgint}
  m (\varphi) = m_0 e^{- \int g_0 \ud \varphi} \,,
\end{equation}
and, at the same time, it allows us to write the relation\footnote{It has to be noted at this point that we are assuming $H-M>0$. Although in principle an arbitrary choice, studies on the causal structure of generalized McVittie \cite{Guariento:2012ri} show that this requirement is the one needed to assure a regular causal structure of the spacetime.}
\begin{equation}\label{f1f2}
  f_2 = -\sqrt{\frac{2}{3}} \left( 6 g_0 \sqrt{-f_1} + \frac{f_1'}{\sqrt{-f_1}} \right) \,.
\end{equation}
This last equality can be seen as another consistency condition that our Lagrangians need to satisfy to allow the generalized McVittie metric to be a solution. We can thus restrict our attention to Lagrangians that satisfy the above, as we did when we imposed conditions \eqref{solg} and \eqref{soll}. Therefore, the class of Lagrangians that satisfy the requirements to admit generalized McVittie metrics as a solution is given by applying Eq.\ \eqref{f1f2} to the action \eqref{action-solution}, that is,
\begin{equation}\label{action-solution-f1f2}
\begin{split}
  S =&\, \int \ud^4 x \sqrt{-g} \left\{ f_1 -\sqrt{\frac{2 X}{3}} \left( 6 g_0 \sqrt{-f_1} + \frac{f_1'}{\sqrt{-f_1}} \right) \vphantom{\frac{1}{2}}\right.\\
  &\left. + 2 X \left[ (2 -\ln X) g_0' - 3 g_0^2 \right] + g_0 \ln X \, \square \varphi + \frac{1}{2} R \right\} \,,
\end{split}
\end{equation}

Finally, we have
\begin{equation}\label{Hdephi}
  H = \sqrt{\frac{-f_1}{3}} - g_0 \dot{\varphi} \,,
\end{equation}
which we recognize as analogous to the algebraic relation connecting the expansion function $H$ and the cuscuton field via its potential [in fact, $f_1 (\varphi)$ plays the role of $-V (\varphi) \propto \varphi^2$ as we will see later].

Notice that once the constraint \eqref{Hdephi} is taken into account when solving for the action \eqref{action-solution-f1f2}, for given $f_1$ and $g_0$ there is only one free function remaining in the solution, which can be either $m$ or $H$, since $\varphi$ can always be redefined. This hints to the presence of a connection between the system given by the Horndeski Lagrangian, that allows the generalized McVittie as solution, and the quadratic cuscuton system studied in \cite{Abdalla:2013ara} that supports the McVittie spacetime. We dub this connection ``cuscuton duality'' and present it in detail in the following section.

\subsection{Cuscuton duality}

Recently, it has been shown that Horndeski fields admit dual representations as different Horndeski fields by the use of disformal transformations \cite{deRham:2014lqa,Zumalacarregui:2013pma,*Zumalacarregui:2012us,Bettoni:2013diz}. In this section, we would like to make use of this idea and analyze a disformal transformation that connects the cuscuton and the Horndeski system studied in the previous sections. More precisely, we show that shear-free solutions of the cuscuton system are transformed into solutions of a dual Horndeski system of the kind examined here for a certain choice of free functions.
 
For simplicity, we work in the Arnowitt-Deser-Misner (ADM) formalism, choosing a foliation of spacetime on which the field is homogeneous and imposing a zero-shift condition $N^i=0$ for the metric. We then start from the cuscuton action \cite{Afshordi:2006ad,*Afshordi:2007yx,Afshordi:2009tt}, which in the ADM formalism reads\footnote{Throughout this section we will perform integrations by parts and ignore the surface terms that appear as a consequence, since they play no role in reaching the conclusions we are trying to reach here anyway.} 
\begin{equation}\label{Scusc-ADM}
  \begin{split}
    S_{\text{c}} =&\, \int \ud t \, \ud^3 x N \sqrt{\gamma} \left[ f_{1c} - \frac{f_{2c}}{\sqrt{2}} \frac{\dot{\varphi}}{ N} \right.\\
    &\left. + \frac{1}{2} \left( \tensor[^{(3)}]{R}{} + K^{a b} K_{a b} - K^2 \right) \right] \,,
  \end{split}
\end{equation}
where $N\equiv\sqrt{-g_{00}}$ is the lapse function, $\gamma_{ij}$ is the 3-metric on each spacelike hypersurface defined in \eqref{3metric}, and $\gamma\equiv \det \gamma_{ij}$ its determinant. As previously stated, the vanishing shift vector, $N^i = 0$, implies that the 3-dimensional Ricci scalar $\tensor[^{(3)}]{R}{}$ is built with the 3-metric only (and its spatial derivatives). $K_{i j}$ is the extrinsic curvature and $K = \tensor{K}{^a _a}$, and they take the form\footnote{We adopt the sign convention for the extrinsic curvature as in \cite{wald}, opposite to the one used in \cite{misner-thorne-wheeler}.}
\begin{align}
  K_{ij} =&\, \frac{1}{N} \dot{\gamma}_{ij} \\
  K =&\, \frac{1}{N} \frac{\dot{\gamma}}{\gamma} \,.
\end{align}
In order to write the action for the cuscuton in the form \eqref{Scusc-ADM} we have also assumed the field $\varphi$ monotonic, in particular satisfying $\dot{\varphi}<0$. For the moment we can leave the function $f_{1c}(\varphi)$ unconstrained, as well as the constant $f_{2c}$, but in order to link to the results discussed in \cite{Abdalla:2013ara} we will ultimately choose
\begin{equation}\label{f-cuscuton}
  \begin{split}
    f_{2c} =&\, \sqrt{2} \mu^2 \\
    f_{1c} =&\, -V(\varphi) = -\frac{3}{4}\mu^4 \left( \varphi(t) + \varphi_0 \right)^2 \,.
  \end{split}
\end{equation}
We apply then the following disformal transformation defined by
\begin{equation} \label{disf-tr-cov}
  g_{\mu \nu} = \alpha (\varphi) \tilde{g}_{\mu \nu} - \left[ \beta (\varphi) - \alpha (\varphi) \right] \frac{\nabla_{\mu} \varphi \nabla_{\nu} \varphi}{2 X} \,,
\end{equation}
such that the lapse and 3-metric transform as
\begin{equation}\label{disf-tr-ADM}
  \begin{split}
    N =&\, \sqrt{\beta} \tilde{N} \,,\\
    \gamma_{i j} =&\, \alpha \tilde{\gamma}_{i j} \,.
  \end{split}
\end{equation}
The extrinsic curvature then transforms as
\begin{equation}
  \begin{split}
      K_{i j} =&\, \frac{1}{2 N} \dot{\gamma}_{i j}\\
      =&\, \frac{\alpha}{\sqrt{\beta}} \left( \frac{1}{2} \frac{\alpha'}{\alpha} \frac{\dot{\varphi}}{\tilde{N}} \tilde{\gamma}_{i j} + \tilde{K}_{i j} \right) \,,
  \end{split}
\end{equation}
while the 3-Ricci scalar transforms like the inverse metric, $\tensor[^{(3)}]{R}{} = \frac{1}{\alpha} \tensor[^{(3)}]{\tilde{R}}{}$.

Under such a transformation the action \eqref{Scusc-ADM} becomes
\begin{equation}
  \begin{split}
    S_{c} =&\, \int \ud t \ud^3 x \tilde{N} \sqrt{\tilde{\gamma}} \sqrt{\frac{ \alpha^{3}}{\beta}}  \left\{  \beta \left[ f_{1c} (\varphi) - \sqrt{\beta} \frac{f_{2c} }{\sqrt{2}} \frac{\dot{\varphi} (t)}{\tilde{N}} \right] \vphantom{\left(\frac{\dot{\varphi}}{\tilde{N}}\right)^2} \right. \\ 
    &+ \frac{1}{2} \left( \frac{\beta}{ \alpha} \tensor[^{(3)}]{\tilde{R}}{} +\tilde{K}^{ab}\tilde{K}_{ab} - \tilde{K}^2 \right) \\
    & \left. + \left[ - \frac{\alpha'}{\alpha} \frac{\dot{\varphi}}{\tilde{N}} \tilde{K} - \frac{3}{4} \left(\frac{\alpha'}{\alpha}\right)^2 \left(\frac{\dot{\varphi}}{\tilde{N}}\right)^2 \right] \right\} \,.
  \end{split}
\end{equation}
Knowing where we are aiming at, we set the functions $\alpha$ and $\beta$ defining the disformal transformation to
\begin{equation}\label{add-tr-def}
  \begin{split}
    \beta (\varphi) \equiv&\, \alpha(\varphi)^3 \\
    \alpha (\varphi) \equiv&\, e^{2 g_0 \varphi(t)} \,,
  \end{split}
\end{equation}
where $g_0$ is a constant that eventually will take the place of the free function $g_0(\varphi)$ appearing in the Horndeski action \eqref{action-solution}. Eqs.\ \eqref{add-tr-def} may also be seen as a time-dependent Lifshitz rescaling of spacetime with $z = 3$. The action then takes its final form
\begin{equation}\label{transf-Scusc}
  \begin{split}
    S_c =& \int \ud t \ud^3 x \tilde{N} \sqrt{\tilde{\gamma}} \left\{ \alpha^{3} \left[ f_{1c}  - \frac{f_{2c} }{\sqrt{2}} \frac{\dot{\varphi}(t)}{\sqrt{\alpha^3}\tilde{N}} \right] - 2g_0 \frac{\dot{\varphi}}{\tilde{N}} \tilde{K} \right.  \\ 
    & \left. - 3 g_0^2 \left(\frac{\dot{\varphi}}{\tilde{N}}\right)^2 +  \frac{1}{2}\left( \alpha^2 \tensor[^{(3)}]{\tilde{R}}{} +\tilde{K}^{ab}\tilde{K}_{ab} - \tilde{K}^2 \right)  \right\} \,.
  \end{split}
\end{equation}

We can now take a look at the Horndeski action \eqref{action-solution} in $3 + 1$ formalism, in the case in which the function $g_0(\varphi) = g_0$ is set to a constant (which can be done, generically, with a field re-definition). We have
\begin{equation}\label{SHorn-ADM}
  \begin{split}
    S_{H} =& \int \ud t \ud^3 x N \sqrt{\gamma} \left[ f_{1H} - \frac{f_{2H}}{\sqrt{2}} \frac{\dot{\varphi}}{N} - 2 g_0 \frac{\dot{\varphi}}{N} K \right. \\
    & \left. - 3 g_0^2 \left( \frac{\dot{\varphi}}{N} \right)^2 + \frac{1}{2} \left( \tensor[^{(3)}]{R}{} + K^{ab}K_{ab} -K^2 \right) \right] \,.
  \end{split}
\end{equation}
Beside the rescaling of the functions $f_{1H}$ and $f_{2H}$, we note a fundamental difference between the two actions: the extra factor $\alpha^2$ multiplying the scalar $3$-curvature in \eqref{transf-Scusc}. Since we are trying to connect the generalized McVittie solution of the Horndeski action to a transformed cuscuton solution, we also want to impose the relationship defining $f_{2H}$ in terms of $f_{1H}$ on the transformed equivalent objects, namely
\begin{subequations}\label{fcfH}
\begin{align}
  f_{1c} =&\, \alpha^{-3} f_{1H} \,, \\
  \begin{split}    
    f_{2c} =&\, \alpha^{-\frac{3}{2}} f_{2H} \\
    =&\, - \alpha^{-\frac{3}{2}} \sqrt{\frac{2}{3}} \left( 6 g_0 \sqrt{-f_{1H}} + \frac{f_{1H}'}{\sqrt{-f_{1H}}} \right) \\
    =&\, - \sqrt{\frac{2}{3}}  \frac{f_{1c}'}{\sqrt{-f_{1c}}} \,,
  \end{split}
\end{align}
\end{subequations}
satisfied by \eqref{f-cuscuton}.

To summarize, so far we have found that transforming the action for the quadratic cuscuton \eqref{Scusc-ADM} via the disformal transformation \eqref{disf-tr-ADM} [with the additional conditions \eqref{add-tr-def} and \eqref{fcfH}] we obtain an action \eqref{transf-Scusc} which takes a form very close to the action for the Horndeski field we originally considered \eqref{SHorn-ADM} when a constant $g_0(\varphi)$ is considered. 

For the disformal transformation we used to be a form of duality between the two systems, we need it to map solutions of one into solutions of the other. Due to the presence of the extra $\alpha^2$ factor in front of the $3$-scalar curvature in the transformed cuscuton action we obtain two conditions that solutions need to satisfy to belong at the same time to the set of classical solutions extremizing \eqref{transf-Scusc} and \eqref{SHorn-ADM}. These are
\begin{equation}\label{anoa-cond}
  \begin{split}
    \tensor[^{(3)}]{R}{}\big|_\text{on shell} =&\, 0 \,, \\
    \tensor[^{(3)}]{R}{_{ij}}\big|_{\text{on shell}} =&\, \frac{1}{N} D_i D_j (N) \,,
  \end{split}
\end{equation}
where $D_a$ is the spatial covariant derivative compatible with $\gamma_{ij}$, defined on each hypersurface of the spacetime foliation. These conditions can be obtained by comparing the equations for the lapse function $N$ and the $3$-metric $\gamma_{ij}$ obtained from the two actions.

We can conclude then that solutions of the cuscuton system \eqref{Scusc-ADM} [with $f_{2c}$ given by \eqref{fcfH}] that satisfy \eqref{anoa-cond} are transformed by the disformal transformation \eqref{disf-tr-ADM} into solutions of the system described by \eqref{SHorn-ADM} where $g_0$ is a constant and $f_{2H}$ is given by \eqref{f1f2}.

The McVittie/quadratic cuscuton solution belongs to this class and is in fact dual to the generalized McVittie solution discussed in the previous section. Applying the disformal transformation to it one obtains
\begin{equation}
  \begin{split}
    \tilde{N} =&\, \alpha^{-\frac{3}{2}}\frac{1-\frac{m_0}{2a(t) r}}{1+\frac{m_0}{2a(t) r}} \\
    =&\, \alpha^{-\frac{3}{2}}\frac{1-\frac{m(t)}{2\tilde{a}(t) r}}{1+\frac{m(t)}{2\tilde{a}(t) r}} \,,
  \end{split}
\end{equation}
as well as
\begin{equation}
  \begin{split}
    \tilde{\gamma}_{i j} =&\, \frac{1}{\alpha} a^2 (t) \left[ 1 + \frac{m_0}{2 a (t) r} \right]^4 g^S_{i j} \\
    =&\, \frac{1}{\alpha} \left[ \frac{m_0}{m(t)} \right]^2 \tilde{a}^2 (t) \left[ 1 + \frac{m(t)}{2 \tilde{a} (t) r} \right]^4 g^S_{i j} \,,
  \end{split}
\end{equation}
where $g^S_{ij}$ is the $3$-dimensional Euclidean metric in spherical coordinates. In the second line of the above equations we have redefined the scale factor as $a(t) = \tilde{a} \frac{m_0}{m(t)}$ in order to get to the usual form of the generalized McVittie metric \eqref{mcvittie}. Here, $m(t)$ is defined as
\begin{equation}
  \frac{m(t)}{m_0} = e^{-g_0 \varphi(t)} \,,
\end{equation}
which is of course consistent with the equations for the Horndeski field. To conclude the transformation, we need a rescaling of the time coordinate
\begin{equation}
  e^{-3g_0 \varphi(t)} \ud t = \ud \tau \,,
\end{equation}
that takes care of the inconvenient $\alpha$ factor in $\tilde{N}$.

Since we started from a specific choice of potential for the cuscuton, and therefore a specific form of its dual $f_{1H}$, the equation \eqref{mdgint} now imposes that 
\begin{equation}\label{mass-cond}
  \frac{1}{g_0} \log \left[ \frac{m_0}{m (\tau)} \right] = \left[ \frac{m(\tau)}{m_0} \right]^3 \frac{2}{\mu^2} \left[ \tilde{H} (\tau) - M (\tau) \right] \,.
\end{equation}
The quadratic cuscuton with a McVittie metric described by the parameter $m_0$ and the expansion function $H(t)$ is then dual to a system consisting of action \eqref{transf-Scusc} which supports a generalized McVittie spacetime described by the dual expansion history $\tilde{H}(\tau)$ and for which the mass function solves \eqref{mass-cond}.

To conclude, we can also notice that the conditions we required for the duality to hold can be summarized by: the choice of functions appearing in the cuscuton side, namely given by the last line of \eqref{fcfH} for $f_{2c}$; the equivalent of it on the Horndeski side \eqref{f1f2}; and the requirement that solutions be spatially flat and satisfy \eqref{anoa-cond}. Although we tailored the transformation to connect the cuscuton/McVittie and Horndeski/generalized McVittie systems, any solution satisfying the above conditions would have a dual via the disformal transformation. In particular, the original cosmological solution for the cuscuton \cite{Afshordi:2007yx} also satisfies the duality requirements and transforms into a rescaled FLRW with the scalar field still defined by the expansion function, even though now in a slightly more complicated way.

\section{Conclusions}\label{sec:conclusions}

The generalization of the McVittie metric proposed by Faraoni and Jacques \cite{Faraoni:2007es} constitutes a fully dynamical exact solution of general relativity that describes a black hole interacting with a surrounding self-gravitating imperfect fluid in an expanding universe. The central mass dependence with time is directly associated with deviations from perfect fluid in the source, which in the comoving frame take the form of a heat flow. This fluid analogy can be made in scalar fields, which always take the form of perfect fluids if the action is of a $k$-essence form, but can acquire dissipative (imperfect-fluid) terms when a more general action is considered. In this work we have considered a subclass of the Horndeski action as a scalar field source minimally coupled to general relativity, and we have shown that the generalized McVittie metric is in fact an exact solution of this system for a specific choice of free functions defining the action, taking the form \eqref{action-solution} where $f_2$ is given by \eqref{f1f2}.

In particular, the heat-flow component of the energy-momentum tensor added by the kinetic gravity braiding term has the correct form to account for the time variation of the central mass. By considering the system formed by the Einstein equations and the field equation,  we have shown that the dynamics of this system reduces to two constraint equations and a first-order differential equation relating the expansion rate with the time-dependence of the field. The equations for the scalar are identically satisfied once these choices are made, and the field dynamics remains unconstrained.

Finally, we have shown that the lack of a differential equation for the evolution of the scalar field is reminiscent of its $k$-essence counterpart: the cuscuton. In fact, the generalized McVittie solution for the action presented here has been shown to be dual to the McVittie/cuscuton via a disformal transformation. Incidentally, it is worth noticing that this disformal transformation coincides with a time-dependent Lifshitz re-scaling of spacetime, with $z = 3$.

The McVittie metric is just an example of a large class of solutions which can be constructed with this technique, and even by itself it provides a great variety of highly nontrivial behaviors. It is remarkable that this special class of non-dynamical fields, like the cuscuton and its disformal dual, admits exact time-dependent solutions with such physically rich structures. Looking into the future, similar to other exact solutions in theoretical physics, we anticipate that this generalized McVittie/Horndeski exact solution could serve as a launching point for studies of novel phenomena in the interface of strong gravity and cosmology. A particularly timely direction could be the phenomenological studies of interaction between dark energy/modified gravity models (many of which can be approximated by the Horndeski action), and astrophysical black holes. 

\begin{acknowledgments}

We thank E.\ Babichev, K.\ Copsey, A.\ M.\ da Silva, C.\ Deffayet, C.\ de~Rham, N. Doroud, G.\ Esposito-Farese, V.\ Faraoni, R.\ Gregory, K.\ Hinterbichler, N.\ Kaloper, B.\ Le Floch, E.\ Papantonopoulos, J.\ Sakstein, V.\ Sivanesan, R.\ Sorkin, T.\ P.\ Sotiriou, D.\ Stojkovic, A.\ J.\ Tolley and A.\ Vikman for very interesting discussions and suggestions. D.\ C.\ G.\ is supported by FAPESP Grants No.\ 2010/08267-8 and No.\ 2013/01854-3. This research was supported in part by Perimeter Institute for Theoretical Physics. Research at Perimeter Institute is supported by the Government of Canada through Industry Canada and by the Province of Ontario through the Ministry of Economic Development \& Innovation.

\end{acknowledgments}

\appendix

\section{Null geodesics in generalized McVittie} \label{null-geodesics}

In the bulk of the McVittie spacetime both horizons are anti-trapping surfaces. However, if at time infinity the metric becomes Schwarzschild-de~Sitter, the inner apparent horizon at time infinity, which we denote by $r_{\infty}$, becomes an event horizon, that is, a null surface which time-like curves starting at some initial event in the bulk can reach and traverse after a finite proper time, entering a region that can no longer causally affect the McVittie bulk \cite{Kaloper:2010ec,Lake:2011ni}.

In the generalized case with time-dependent mass, one needs to carry out an analog analysis. As in McVittie with constant mass, the analysis of whether metric \eqref{mcvittie} with $m (t)$ consists of a black hole is centered on the fate of radial ingoing null geodesics. If, after departing from some initial radius $r_0$ an ingoing null geodesic reaches the surface $r_{\infty}$ (defined as the position of the inner horizon at time infinity) after a finite affine parameter $\lambda$ has elapsed, then the spacetime is geodesically incomplete, and therefore $r_\infty$ is an event horizon\footnote{Recall that, as in the McVittie case, it is sufficient for the curvature to be well behaved on $r_{\infty}$ that $H_0 > 0$ and $M = 0$ at time infinity \cite{Guariento:2012ri}.}. On the other hand, if it takes an infinite affine parameter interval for the geodesic to reach $r_\infty$, then this surface consists of a null infinity, and therefore the metric is geodesically complete, and not a black hole.

In areal radius coordinates, the generalized McVittie metric reads \cite{Guariento:2012ri}
\begin{equation}
  \ud s^2 = - R^2 \ud t^2 + \left[ \frac{\ud \hat{r}}{R} - \left( H - M + \frac{M}{R} \right) \hat{r} \ud t \right]^2 + \hat{r}^2 \ud \Omega^2 \,,
\end{equation}
with the areal radius defined as $\hat{r} \equiv a r \left( 1 + \frac{m}{2 a r} \right)^2$ and the functions $R \equiv \sqrt{1 - \frac{2 m}{\hat{r}}}$, $M \equiv \nicefrac{\dot{m}}{m}$ and $H \equiv \nicefrac{\dot{a}}{a}$. The null radial geodesic equation for the $t$ coordinate reads
\begin{equation}
\frac{\ud^2 t}{\ud \lambda^2} = - \left[ M + \frac{H - M}{2} \left( R + \frac{1}{R} \right) - \frac{2 m}{\hat{r}^2} \right] \left( \frac{\ud t}{\ud \lambda} \right)^2 \,.
\end{equation}

In the original analysis, with $M = 0$ \cite{Lake:2011ni}, the coefficient of $\left( \frac{\ud t}{\ud \lambda} \right)^2$ is shown to be always larger than the function $h (\hat{r})$ defined as
\begin{equation}
  h (\hat{r}) \equiv \frac{2 m}{\hat{r}^2} - \frac{R}{2 \hat{r}} \left( R + \frac{1}{R} \right) \,,
\end{equation}
so that the geodesic equation satisfies the inequality
\begin{equation}\label{ineq-geod}
  \frac{\ud^2 t}{\ud \lambda^2} \geq h (\hat{r}) \left( \frac{\ud t}{\ud \lambda} \right)^2 \,,
\end{equation}
which can be solved analytically and is shown to diverge for $t \to \infty$, thus showing that radial ingoing null geodesics reach the surface $r_- (\hat{r}, t \to \infty)$ after a finite affine parameter $\lambda$. In our case, we need to take into account the presence of a nonzero $M$. Between the two apparent horizons, the function
\begin{equation}
  f(\hat{r}, t) \equiv R^2 - \hat{r}^2 \left( H - M + \frac{M}{R} \right)^2
\end{equation}
is positive. For $M > 0$, this can be rewritten as
\begin{equation}
  H - M + \frac{M}{R} < \frac{R}{\hat{r}} \,,
\end{equation}
which may be cast as
\begin{equation}
  \frac{H - M}{2} \left(R + \frac{1}{R} \right) + M < \frac{R}{2 \hat{r}} \left( R + \frac{1}{R} \right) - \frac{M}{2 R^2} \left( 1 - R^2 \right) \,.
\end{equation}
This implies
\begin{equation}
\begin{split}
  \frac{2 m}{\hat{r}^2} &- \frac{H - M}{2} \left(R + \frac{1}{R} \right) - M \\
  &> \frac{2 m}{\hat{r}^2} - \frac{R}{2 \hat{r}} \left( R + \frac{1}{R} \right) + \frac{\dot{m}}{\hat{r} - 2 m}\\
  &> h \,.
\end{split}
\end{equation}
so Eq.~\eqref{ineq-geod} still holds for $M > 0$. This means that generalized McVittie metrics with accreting mass functions are geodesically incomplete between the two apparent horizons, and therefore that radial ingoing null geodesics reach the surface $(\hat{r}_-, t \to \infty)$ in finite proper time.

\bibliography{shortnames,referencias}

\end{document}